# Meteor Science

## A practical method for the analysis of meteor spectra

*Martin Dubs*[1] *and Peter Schlatter*[2]

The analysis of meteor spectra (photographic, CCD or video recording) is complicated by the fact that spectra obtained with objective gratings are curved and have a nonlinear dispersion. In this paper it is shown that with a simple image transformation the spectra can be linearized in such a way that individual spectra over the whole image plane are parallel and have a constant, linear dispersion. This simplifies the identification and measurement of meteor spectral lines. A practical method is given to determine the required image transformation.

## 1 Introduction

Meteor spectra are recorded by placing a prism or a transmission grating in front of the camera lens (Rendtel, 2002). The light of any point source is separated into a line spectrum, with different wavelengths refracted or diffracted at different angles. In prisms, the wavelength separating mechanism is refraction, which is highly nonlinear. This is described by the dispersion $\mathrm{d}\beta/\mathrm{d}\lambda$, the change in refraction angle per wavelength unit. It is a strongly varying function of the wavelength $\lambda$ and depends on the prism angle and prism material. In gratings the separation of different wavelengths is caused by diffraction from the closely spaced grating lines and dispersion is a slowly varying function of the incident and exit angles and inversely proportional to the separation of grating lines.

There are other differences between prisms and gratings. Prism angular dispersion is generally small, requiring long focal lengths for sufficient linear dispersion. An advantage of prisms is that all light is separated into one spectrum. Gratings on the other hand produce spectra of different orders with different dispersion. Part of the light passes through the grating undiffracted (the so called zero order), which is used as a zero wavelength reference. If recorded, this is of great help for the calibration of the spectrum. Modern gratings are blazed, that means that most of the light is diffracted into one (often the first) order with an efficiency of typically 50% or higher. The rest of the light produces the zero order and other (higher) orders on both sides of the zero order.

In this paper only grating spectra are discussed, as they are at present the preferred choice for video and CCD cameras with a small chip size (compared to large size photographic film). Unfortunately the same method cannot be applied to prism spectra, where the nonlinearities are much greater and of a different origin.

For a given chip size the focal length of the lens determines the field of view and for a given grating also the linear dispersion $\mathrm{d}x/\mathrm{d}\lambda$ in $\mu$m/nm or pixel/nm. As the light of a meteor is dispersed over many pixels the detection sensitivity is several magnitudes lower for meteor spectra than for the detection of meteors with the same lens detector combination, so fewer meteors are recorded. Choosing a short focal length increases the field of view but reduces the linear dispersion or resolution of the spectrum. In addition, at larger incident and diffracted angles the nonlinearity of dispersion becomes more apparent, making the analysis of spectra quite complicated. Both, the low number of useful events and the complicated analysis of the spectra discourages many observers of recording meteor spectra.

In this paper, the calibration of meteor spectra is treated in some detail. Based on simple geometric analysis a practical method is given which straightens the curved, nonlinear spectra to parallel, linear spectra with constant dispersion over the whole field of view. The geometric approach also suggests a method for determining the required image transformation, which will be discussed in detail below.

## 2 Theory of grating diffraction

A method for computing the wavelengths of objective grating spectra, suitable for analysis of meteor spectra, is described in (Ceplecha, 1961). At the time the Ceplecha paper had been written, computers were not in widespread use and photographic film was the recording medium. Today software for image analysis and CCDs are in common use and the analysis of the spectra should take advantage of the increased possibilities. For easier comparison with that work, the same notation and coordinate systems as far as convenient will be used in the present paper.

Figure 1 shows the orientation of a Cartesian coordinate system with respect to the grating. The plane of the grating coincides with the $xy$-plane and the grooves are aligned parallel to the $y$-axis. Light from a meteor trail can be regarded as a succession of point sources at infinity, each point being characterized by parallel rays that eventually impinge on the grating. In the given coordinate system, the components of a unit vector $(A, B, C)$ describe the direction of the rays originating from one point, while the components of the unit vector $(A', B', C')$ describe the direction of the diffracted

[1]Im untern Stieg 2, CH-7304 Maienfeld, Switzerland.
Email: martin-dubs@bluewin.ch
[2]Birkenweg 8, CH-3033 Wohlen b. Bern, Switzerland

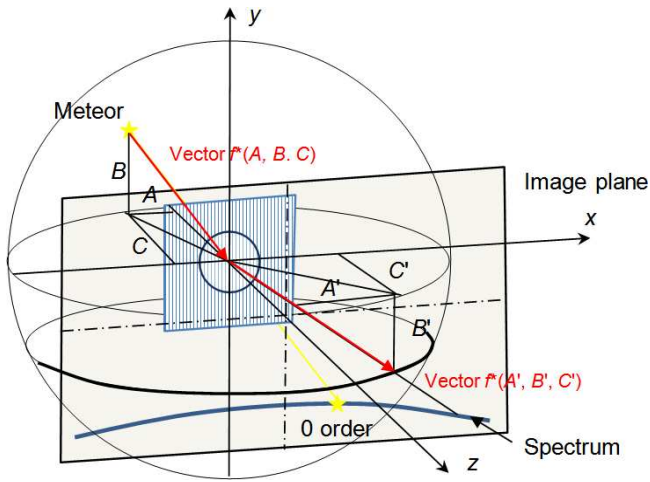

Figure 1 – Coordinate system. $(A, B, C)$ components of incident ray unit vector. $(A', B', C')$: components of diffracted ray unit vector, projected on a sphere with radius $f$.

beam. The grating equation relates these two vectors (Rowland, 1893):

$$A' = A + m\lambda G \quad (1)$$

$$B' = B \quad (2)$$

$$C' = \sqrt{(1 - A'^2 - B'^2)} \quad (3)$$

$\lambda$ denotes the wavelength of the incident beam and $m$ is the grating order. The special case $m = 0$ is called zero order, for which the incident beam is not deflected, independent of the wavelength. $G$ is the grating constant or inverse grating line-spacing in grooves/mm.

In textbooks on optics, the grating equation is usually given in angular notation, see e.g. (Schroeder, 1970):

$$m\lambda G = \cos\gamma(\sin\beta + \sin\alpha) \quad (4)$$

with $\alpha$ denoting the angle of incidence, $\beta$ the angle of diffraction and $\gamma$ the angle between the incident ray and the $xz$-plane. While the angular notation is equivalent to the vector notation of equations (1–3), the vector notation considerably facilitates the subsequent derivations.

## 3 A basic lens model

The orientation of the meteor camera relative to the coordinate system is shown in Figure 2. The optical axis of the lens is coincident with the $z$-axis and both the image plane and the grating plane are at right angles to the $z$-axis. In order to find the image point $P$ of the diffracted ray $(A', B', C')$, a basic model of the lens is required. If we assume that the lens is free of aberrations except distortion, there is rotational symmetry with respect to the optical axis. Then, the distance $r$ of an image point $P$ from the optical axis is entirely determined by the angle between the diffracted ray and the $z$-axis, the polar angle $\rho$:

$$r = fg(\rho) \quad (5)$$

where $f$ denotes the focal length of the lens. The function $g(\rho)$ determines the projection properties of the lens. It must be an odd, strictly increasing function, and for small angles of $\rho$, i.e. for paraxial rays, $g(\rho) = \rho$. A Taylor expansion will consist of terms with odd exponents only. An ideal lens, for example, is characterized by the so called gnomonic projection, for which $g(\rho) = \tan(\rho)$ (Calabretta & Greisen, 2002).

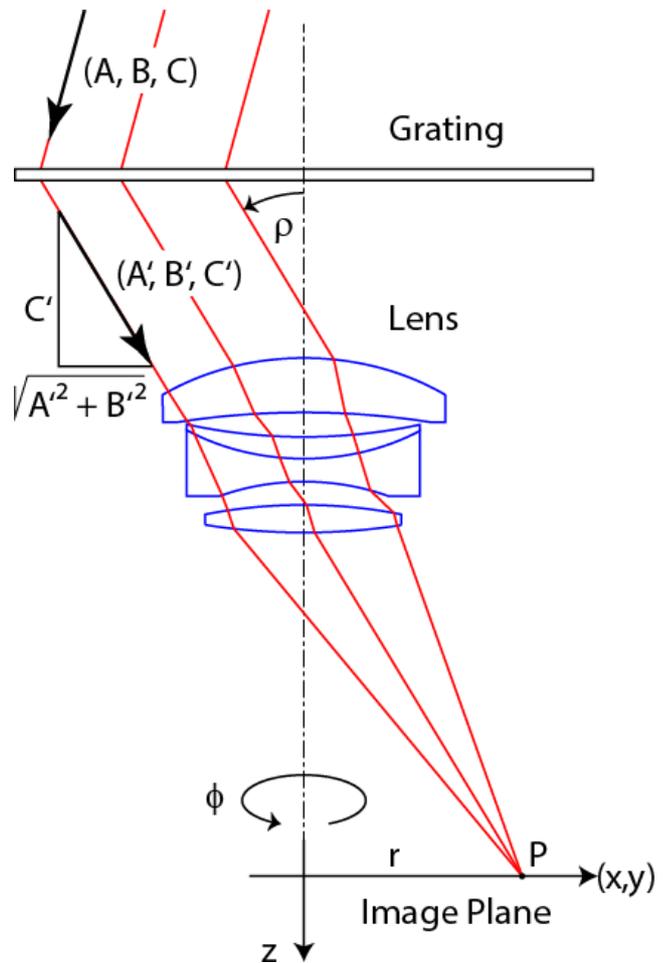

Figure 2 – Section through meteor camera showing relation between $\rho$ and $r$.

With equation (5), the coordinates of the image point $P$ are given by

$$x = r\cos\phi = fg(\rho)\cos\phi \quad (6a)$$

$$y = r\sin\phi = fg(\rho)\sin\phi \quad (6b)$$

where $\phi$ refers to the azimuth angle of point $P$, measured in the $xy$-plane. Both the azimuth angle $\phi$ and the polar angle $\rho$ may be expressed in terms of the unit vector $(A', B', C')$, see Figure 2:

$$\sin\rho = \sqrt{(A'^2 + B'^2)}$$
$$\cos\phi = A'/\sqrt{(A'^2 + B'^2)} = A'/\sin\rho$$
$$\sin\phi = B'/\sqrt{(A'^2 + B'^2)} = B'/\sin\rho$$

Substituting the last two equations into equations (6a) and (6b) and taking into account equations (1) and (2) leads to

$$x = fg(\rho)/\sin(\rho)(A + m\lambda G) \quad (7a)$$

$$y = fg(\rho)/\sin(\rho)B \quad (7b)$$

These two relations are valid for any lens with projection properties $g(\rho)$. They define how the rays of a point source in the sky at $(A, B)$ are mapped to the sensor at $(x, y)$.

A special choice for $g(\rho)$ is the orthographic projection. It is defined by $g(\rho) = \sin(\rho)$ (Calabretta & Greisen, 2002). Inserting this into equations (7a and 7b), the $\sin(\rho)$ term in the denominator and the explicit dependence of $(x, y)$ on $\rho$ are eliminated:

$$x = f(A + m\lambda G) \quad (8a)$$
$$y = fB, \quad (8b)$$
$$dx/d\lambda = fmG \quad (9)$$

The orthographic projection maintains the linearity of the vector components in equations (1) and (2), and a polychromatic point source is expanded into a spectrum with ideal properties:

- The spectrum extends along a straight line parallel to the $x$-axis.
- The linear dispersion has a constant value over the entire image plane. The dispersion may easily be determined from two known spectral lines or the zero order and one spectral line.
- Individual spectra of the points that make up a meteor trail are shifted in $x$ and in $y$ and are parallel to each other. Since the dispersion has a location-independent value, one global calibration suffices to reduce all spectra.

Unfortunately, lenses do not map objects according to an orthographic projection. Rather an ideal lens is characterized by the gnomonic projection, defined by $g(\rho) = \tan \rho$. Ceplecha's calculations are in fact based on an ideal lens. This leads to curved spectra, the so called "diffraction hyperbola" and to non-linear and location-dependent dispersion relations. But even high-quality lenses show some distortion and deviate from a gnomonic projection, thereby modifying the hyperbola in a complicated way. And particularly wide-angle lenses, which are popular in meteor cameras, are affected by distortion.

We now show that these difficulties can be avoided by applying an image transformation that radially distorts the image in such a way that the resulting projection becomes orthographic. After this transformation, equations (8) and (9) apply. The spectra are rectified and the dispersion gets constant over the entire field.

The required transformation maps a point in the original image, $P = (r, \phi)$, to a point in the radially modified image, $P' = (r', \phi)$. The azimuth angle $\phi$ is left unchanged and, by the definition of the orthographic projection, the transformed radius must satisfy the equation

$$r' = f \sin \rho \quad (10)$$

Inverting the function $g$ in equation (5) and solving for the polar angle $\rho$ leads to the prescription for the transformation:

$$r' = f \sin[g^{-1}(r/f)] \quad (11a)$$

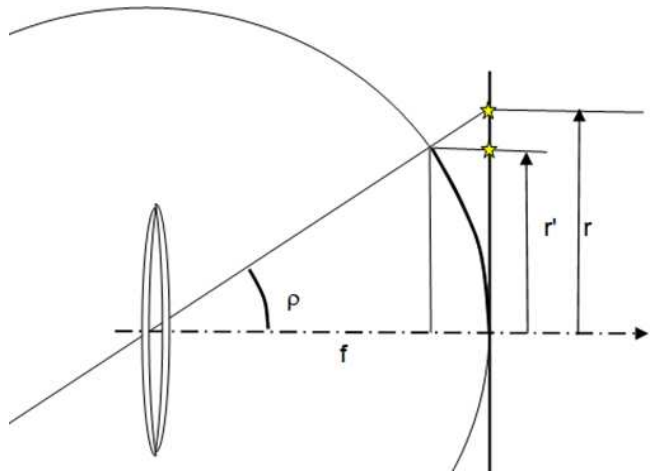

Figure 3 – Relation between the gnomonic projection ($r$) and the orthographic projection ($r'$) of a point on the sphere with radius $f$. The prime denotes the coordinates in the orthographic projection coordinate system.

The inverse function

$$r = fg[\arcsin(r'/f)] \quad (11b)$$

is required for the practical calibration example and for the implementation of the transformation in an image processing software.

For the tangential (gnomonic) projection as a special case (no lens distortion) the transformation to the orthographic projection can be given in explicit form (see also Figure 3):

$$r = f \tan[\arcsin(r'/f)] = r'/\sqrt{[1 - (r'/f)^2]} \quad (12)$$

The function $g(\rho)$, which is required for the transformation, must be determined experimentally for each lens/sensor combination. Several methods may be considered, e.g. an astrometric analysis of a star field or a direct measurement on an optical bench.

As will be shown in a practical example, it is possible to define the transformation without resorting to $g(\rho)$ by directly analyzing a calibration spectrum. This method relies on the fact that both $\sin(\rho)$ and $g(\rho)$ are odd functions. Equation (11b) can be represented by a polynomial with odd exponents in $r/f$:

$$r = f(r'/f + a_3(r'/f)^3 + a_5(r'/f)^5 + \ldots) = r'(1 + a_3(r'/f)^2 + a_5(r'/f)^4 + \ldots) \quad (13)$$

For the tangential projection the polynomial coefficients of equation (12) are given by

$$r = r'(1 + \frac{1}{2}(r'/f)^2 + \frac{3}{8}(r'/f)^4 + \frac{5}{16}(r'/f)^6 + \ldots) \quad (14)$$

## 4 Equipment

Before describing the experiments, an overview of the used equipment may be useful, although the method is applicable to any meteor camera with a grating, if some important details are taken into account (in particular, the grating has to be mounted perpendicularly to the optical axis).

The meteor station at Maienfeld is equipped with two Watec 902H2 ultimate video cameras, one operating for the Swiss Meteor Network and supplying data to the FMA (Fachgruppe Meteorastronomie) database. This has a Computar HG2610AFCS-HSP lens ($f = 2.6$ mm, $f/1.0$) for recording and measuring time resolved meteor tracks. Together with the other stations of the network this gives the information about meteor path, velocity and distance. Without this information, the spectra alone would be much less useful.

The second camera is equipped with a zoom lens (Tamron 12VG412ASIR 1/2", $f$: 4–12 mm, $f/1.2$) and with a blazed 300 l/mm grating (Thorlabs GT50-03, blaze angle 17.5°, 50 × 50 mm).[a]

The grating has been changed recently to a 600 l/mm grating, for which however not many useful results exist yet. The zoom lens is quite convenient. In order to test and optimize the method and capture numerous spectra a short focal length was used. For the analysis of the meteor spectra resolution was not sufficient, so the focal length was increased. With luck some nice spectra were recorded, but still of limited scientific value. The 600 l/mm grating doubles the resolution within the same field of view, producing usable meteor spectra.

## 5 Calibration

In general, neither the exact grating constant nor the focal length and the distortion coefficients are precisely known, so some way of calibration is necessary. In addition, the rotational symmetry of the distortion correction given by the equations (11–14) above is only valid if the grating is mounted perpendicularly to the optical axis, so some attention should be given to verify this. Also the position of the optical axis is generally not exactly in the centre of the detector but a few pixels offset in the $x$- and $y$-coordinate at $(x_0, y_0)$. Notice also that image coordinates are usually measured in pixels, with the origin at a corner. This will be assumed in the following unless otherwise noted. The radius $r = \sqrt{(x - x_0)^2 + (y - y_0)^2}$ is also measured in pixels. The position of the optical axis on the image sensor can be determined before mounting the grating, by imaging the night sky, if necessary with stacking several images and finding an astrometric solution of the coordinates of the stars in the image. If the software UFO Capture[b] is used for the acquisition of the images and UFO Analyzer for the astrometric solution of the images, both the aspect ratio of the pixels and the coordinates of the optical axis are fitted in addition to distortion parameters. If other image acquisition software is used, it may be necessary to calculate them, depending on the form of the astrometric solution. If the software for the radial transformation according to equation (13) assumes square pixels, the image has to be stretched in the $y$-direction by the corresponding aspect ratio factor, otherwise the rotational symmetry is lost and the transformation will not be correct for all parts of the image.

[a]http://www.thorlabs.de/newgrouppage9.cfm?objectgroup_id=1123
[b]http://sonotaco.com/soft/e_index.html

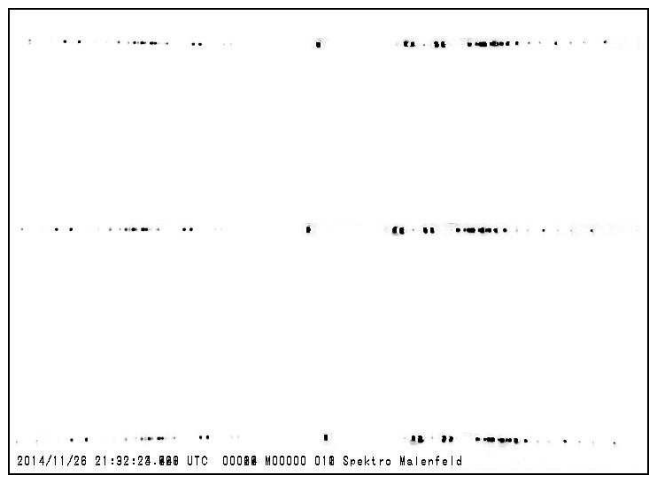

Figure 4 – Composite image of calibration lamp spectrum (Hg-Ar, Ocean Optics, orders −3 to 3) recorded in different parts of the image, superposed into a single image. Curvature is visible for the top and bottom spectra.

### 5.1 Calibration spectra

Depending on the desired accuracy several calibration spectra at different values of $y$ should be recorded, stretching over the full width and height of the detector. In principle it is sufficient to record a calibration spectrum at $y = y_0$ (straight spectrum). For the calibration a suitable light source with lines with known wavelengths is used, which gives spectral lines over the whole width of the detector. For a start a monochromatic laser with known wavelength (e.g. He-Ne gas laser at 632.8 nm or a blue-ray laser at ≈ 405 nm) gives a course calibration of dispersion, in particular for wide angle lenses and/or gratings with low dispersion. Of course, higher order spectral lines should be used for the calibration as well, possibly taken with longer exposure times to see them. Quite useful is also an Hg-Ar calibration lamp[c] with several lines from UV to near-IR.

### 5.2 Calibration example

Figure 4 shows a composite image of a calibration lamp (Hg-Ar, Ocean Optics) recorded in the upper, centre and lower part of the image, combined into a single image. The image was corrected for the non-square pixel shape by a scaling factor of 0.9183 in the $y$-direction (obtained from an astrometric analysis of sky images in UFO Analyzer). The zoom lens was adjusted to approx. 7 mm focal length, the same as used for recording meteors in a longer detection run. Only the centre spectrum was used to determine the dispersion of the grating/lens combination and the transformation parameters for changing the actual image into an orthographic projection. A calibration function was fitted to the measured line positions with the method of least squares. The fitting function was obtained by using equation (8a), replacing $x'$ by $r'$ for $y = y_0$ and inserting $r'$ into equation (12). The parameter $c_2$ was introduced to account for lens distortion by interpolating between a gnomonic projection ($c_2 = (p/f)^2$) and an orthographic projection ($c_2 = 0$). This corrects for 3rd order lens distortion and gives a good approximation for the 5th order term:[d]

[c]http://oceanoptics.com/product/hg-1/

Table 1 – Measured positions of selected (non-overlapping) Hg lines in the central spectrum of Figure 4, together with calibration wavelengths (NIST)[e] in different orders and fitted positions according to equation (15).

| $m \cdot \lambda_{NIST}$ [nm] | $x$ [pixel] | fit $x$ [pixel] | error [pixel] | Line |
|---|---|---|---|---|
| −1307.498 | 13.24 | 13.38 | −0.14 | Hg 3rd order |
| −1092.147 | 68.75 | 68.58 | 0.18 | Hg 2nd order |
| −871.666 | 124.25 | 124.03 | 0.23 | Hg 2nd order |
| −809.313 | 139.37 | 139.54 | −0.17 | Hg 2nd order |
| −546.074 | 204.43 | 204.41 | 0.02 | Hg |
| −404.656 | 238.75 | 238.90 | −0.16 | Hg |
| 0 | 336.80 | 336.80 | −0.00 | zero order |
| 404.656 | 434.22 | 434.38 | −0.16 | Hg |
| 546.074 | 468.83 | 468.62 | 0.21 | Hg |
| 809.313 | 532.87 | 532.77 | 0.09 | Hg 2nd order |
| 871.666 | 548.13 | 548.08 | 0.05 | Hg 2nd order |
| 1092.147 | 602.45 | 602.65 | −0.20 | Hg 2nd order |
| 1307.498 | 656.77 | 656.72 | 0.00 | Hg 3rd order |

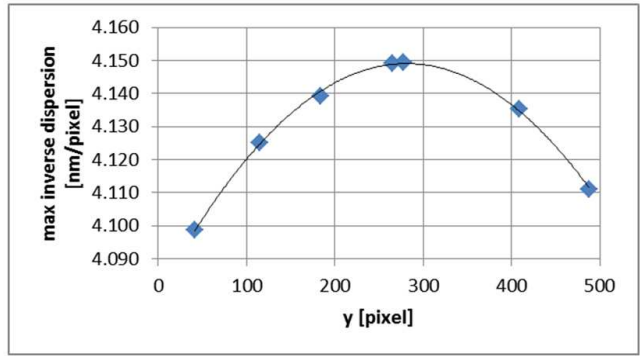

Figure 5 – Inverse dispersion from fit of measured spectra in different parts of the image.

$$x = x_0 + (\lambda - \lambda_0)/disp_0/\sqrt{[1 - c_2((\lambda - \lambda_0)/disp_0)^2]} \quad (15)$$

(($\lambda - \lambda_0)/disp_0$ corresponds to $x' - x_0 = r'$ in the orthographic projection) with the following fit parameters: $disp_0 = (d\lambda/dx)_0 = 4.145$ nm/pixel (inverse dispersion), $x_0 = 362.2$ pixel, $\lambda_0 = 104.8$ nm (offset of $\lambda$ at $x_0$), $c_2 = 4.104 \cdot 10^{-7}$.

The measured line positions together with the fitted positions used for the calibration are shown in Table 1.

From the inverse dispersion $(d\lambda/dx)_0$ and the known grating constant the focal length is calculated from equation (9) as $f = 6.92$ mm.

$x_0$ is the position determined for the symmetry centre of the fit function. Ideally it is located in the image centre, but small deviations may occur if the lens or the grating are not perfectly aligned.

From the fit parameters above, it is possible to calculate the distortion coefficients according to equation (13):

$$r = r'[1 + 2.052 \cdot 10^{-7} r'^2 + 6.318 \cdot 10^{-14} r'^4], \quad (16)$$

with $r$ and $r'$ measured from the apparent centre $(x_0, y_0)$. The value of $y_0$ can be determined from the variation of the dispersion $(d\lambda/dx)_0$ as a function of $y$, by calibrating spectra at different $y$-values and determining its maximum. This is shown in Figure 5, with resulting values of $(x_0, y_0) = (362.2, 281.3)$.

With the position of the symmetry centre and the coefficients of equation (13), the distortion parameters of this lens are known and the transformation can be applied to the image of the spectrum, with the result shown in Figure 6. The slight curvature of the top and bottom spectrum is eliminated and the linearity of the calibration can be checked. With a single inverse dispersion of 4.145 nm/pixel, spectra for different $y$-values can be calibrated with an rms error of 0.94 nm or 0.23 pixel (Figure 7). The error is mostly caused by saturated spectral lines and only slightly larger than the rms error for a 5th order polynomial fit to the dispersion function of a single spectrum before applying the transformation. In addition the data show that in the corners the errors are largest. This is to be expected since the fit was done on the $x$-axis out to a radius of 360 pixel, with the half diagonal of 446 pixel being considerably larger. The fit could be improved by simultaneously fitting several spectra at different $y$ values or by a separate determination of lens distortion. This was not done here.

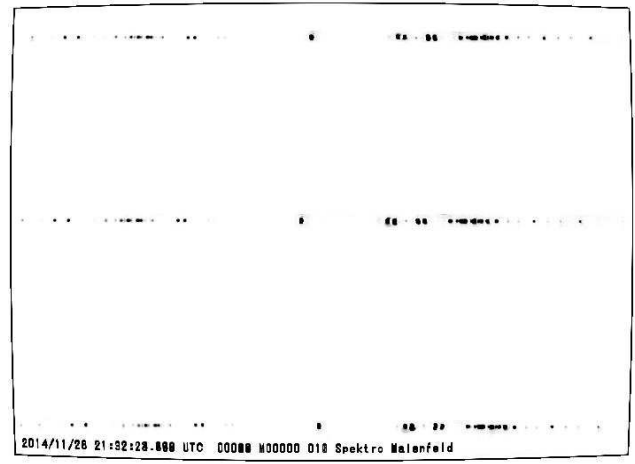

Figure 6 – Image of Figure 4, after applying the image transformation to correct the curvature of the spectrum and nonlinear dispersion.

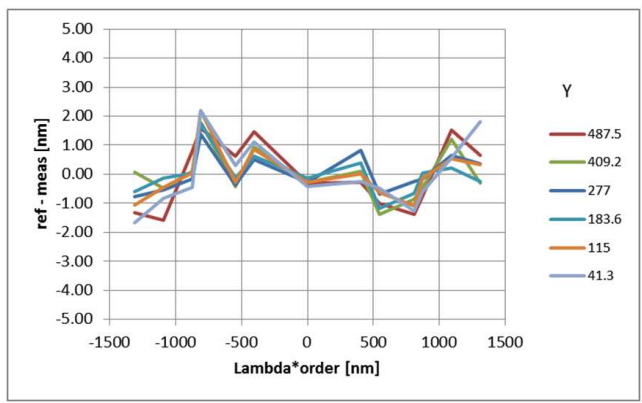

Figure 7 – Wavelength error of measured lines, compared to their computed position, assuming a constant dispersion of 4.145 nm/pixel for all the spectra.

---

[d]The distortion correction with $c_2$ is similar to the fit of lens distortion by the law (Kwon et al., 2014) $r = k_1 \cdot \sin(\beta/k_2)$ with adjustable parameters $k_1, k_2$, which interpolates between orthographic ($k_2 = 1$) and equidistant ($k_2 \to \infty$) projection.

[e]NIST, Atomic spectra database, http://physics.nist.gov/PhysRefData/ASD/lines_form.html

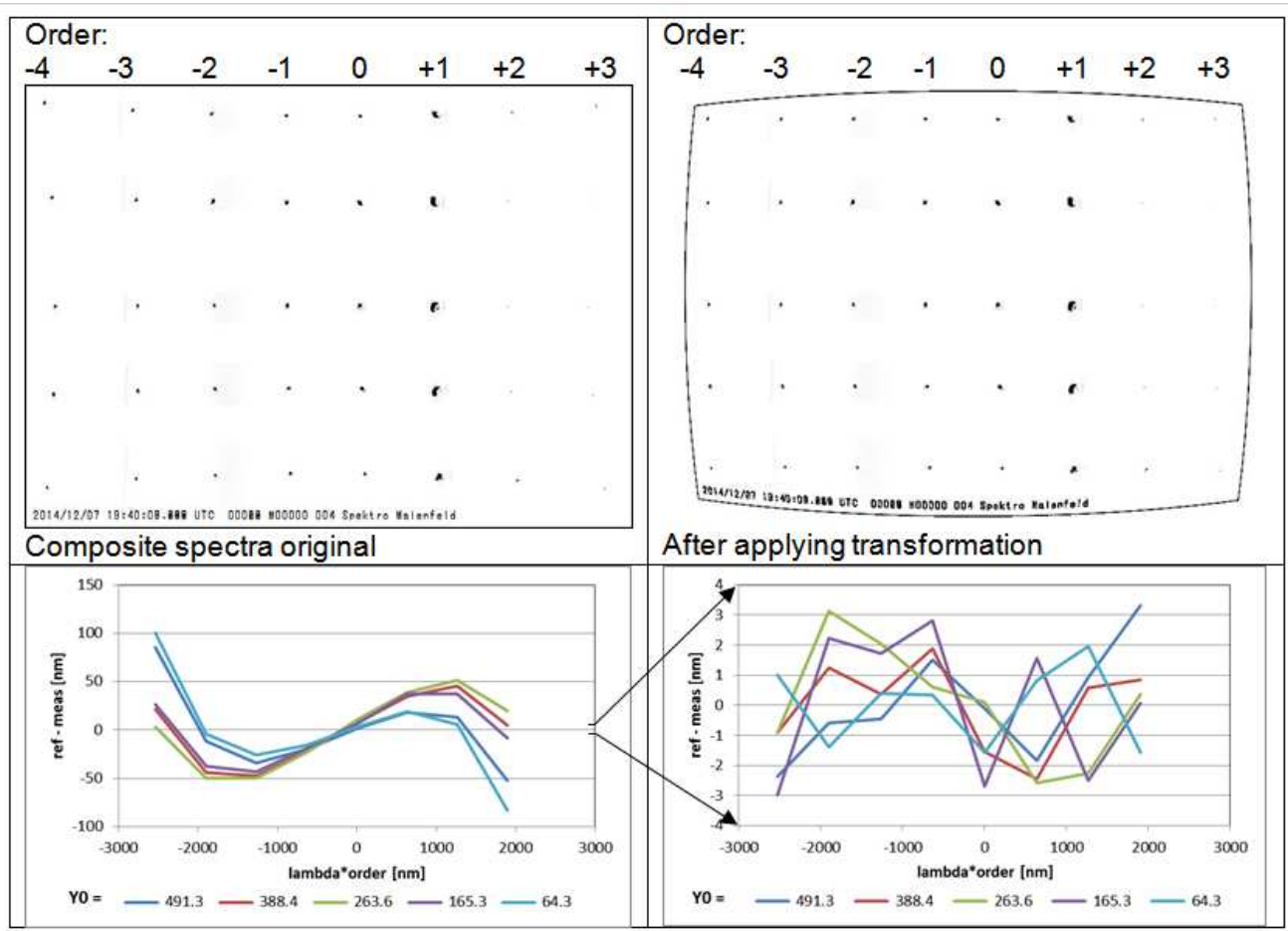

*Figure 8* – Spectra before and after correction of grating nonlinearity and lens distortion. Left: Uncorrected data. Right: After application of the image transformation. Top: Spectra of He-Ne laser. Bottom: Errors from a calibration assuming a linear dispersion law. The blaze of the grating reduces the intensity of the $2^{\text{nd}}$ and third order spectrum to $< 0.5\%$, making them barely visible, but sufficiently bright for analysis, while the first order is overexposed.

### 5.3 $2^{\text{nd}}$ example, short focal lens, not useful for meteor spectroscopy

The following example, although not of practical value for meteor spectroscopy (too low spectral resolution), shows the effect of grating nonlinearity and lens distortion. The images were recorded with the same equipment as the example above, the difference being a shorter focal length (approx. 4 mm) and a He-Ne laser for calibration. At this wide angle enough orders of the He-Ne laser line at 632.8 nm are recorded for a symmetric $5^{\text{th}}$ order polynomial fit of $x$ vs. $\lambda$ around $x_0$. The images before and after correction of distortion are shown in Figure 8. The coefficients for the correction of the distortion were: $(x_0, y_0) = (367.0, 286.5)$, $disp_0 = 7.354$ nm/pixel $\rightarrow f = 3.97$ mm, $r = r'[1 + 3.94 \cdot 10^{-7} r'^2 + 2.01 \cdot 10^{-12} r'^4]$. The rms error of the linear calibration after the transformation was 1.7 nm or 0.24 pixel.

### 6 Analysis of meteor spectra

Once calibrated and without changing grating orientation or focal length of the lens, spectra of meteors can be analysed with the following procedure.

- For video spectra the file is converted to single images. If desired, the video frames of an interlaced video may be deinterlaced into fields with higher time resolution. This results in higher spectral resolution, if the meteor velocity has a component in the direction of the dispersion (along $x$-axis).

- Dark frames are subtracted. A master dark can be obtained by averaging images before or after the appearance of the meteor. This subtraction also eliminates background stars, which otherwise could contaminate the meteor spectra.

- These images are stretched in the $y$-direction if necessary to produce square pixels. Then the transformation to the orthographic projection is applied to all images with meteor spectra. In a streamlined workflow, the image extraction from the video file with or without deinterlacing, the stretching and transformation to the orthographic projection can be combined into a single procedure.

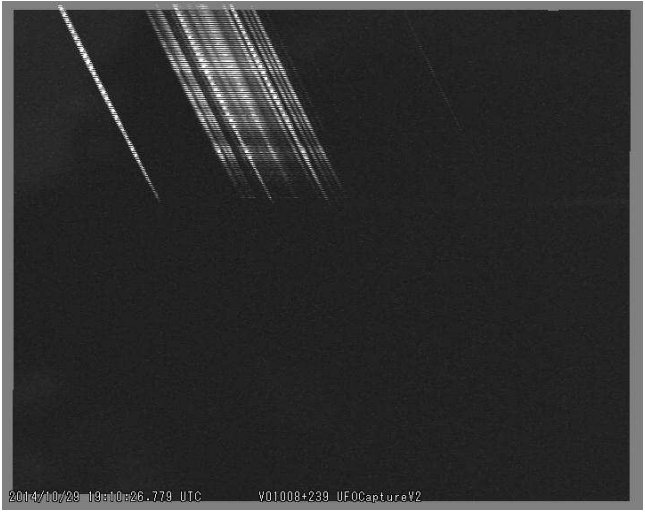

*Figure 9* – Meteor 2014 October 29, $19^{\text{h}}10^{\text{m}}26^{\text{s}}$ UT at Maienfeld, peak image of video, duration of recorded path 0.68 sec, grating 300 l/mm.

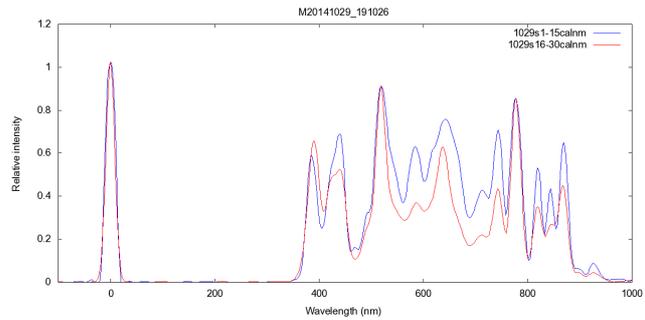

*Figure 10* – Calibrated spectrum, blue curve: $1^{\text{st}}$ part, video fields 54–68 of orthographic projection registered and stacked (average magnitude $-4.7$ mag). Red curve: 2nd part, video fields 69–83 (average magnitude $-3.7$ mag). Strong lines including zero order are broadened by saturation, especially in the first spectrum.

- The meteor spectra, because they are parallel and all have the same dispersion, can be stacked with the zero order as a suitable reference in order to increase signal to noise ratio. In case the zero order is not visible, another prominent line of the spectrum can be used as reference for stacking. Many video cameras only have 8-bit resolution, so averaging several frames is quite important.

- The resulting spectral image is converted to a 1-dimensional raw spectrum and calibrated with the known dispersion from the lamp calibration, using the zero order as a reference. As a check some well-known lines (e.g. Na-D or O I) should show up with the correct wavelength. Minor adjustments of calibration to compensate for a shift of focal length may be applied at this point. In the absence of the zero order (outside of image) the constant linear dispersion of the spectrum helps to identify some known meteor spectral lines and find the wavelength reference position. A correct line assignment of an unknown spectrum with not well known dispersion would be quite difficult.

- If available, the spectrum may be corrected for spectral response obtained from a spectrum of a light source with known spectral energy distribution (calibrated star or tungsten lamp with known blackbody temperature).

### 6.1 Meteor spectrum of 2014 October 29

The meteor of 2014 October 29, $19^{\text{h}}10^{\text{m}}26^{\text{s}}$ UT was selected as an example. It appeared right in the corner of the field of view, so the required distortion correction was the maximum possible (Figure 9). Peak magnitude of the meteor was $-4.9$ mag, with the zero order and the strongest lines overexposed due to the limited dynamic range of the video camera, which reduces the usability for a quantitative analysis of the intensities. The spectra of the individual video frames were processed as described above. After de-interlacing 30 spectra were dark corrected, transformed to the orthographic projection, and registered to align to the zero order. Two series of 15 spectra each were stacked. From these images the 1-dimensional spectra were extracted and calibrated with the zero order and the known inverse dispersion of 4.145 nm/pixel, verified by the O I lines at 777.4 nm in first and second order.

## 7 Conclusion

Meteor spectra do appear anywhere in the field of view, so the assignment of spectral lines may be difficult if the exact dispersion of the spectrum is not known and varies in different areas of the image. Using the described transformation of the spectra to an orthographic projection solves this problem. With the known wavelength of a single spectral line (e.g. the Na-D line), the whole spectrum can be calibrated for any position of the meteor in the image area. This is particularly useful if the zero order is outside the field of view. The determination of the transformation coefficients requires some effort, but it has to be done only once for each lens (at fixed focal length and fixed grating orientation). The transformation to a constant, linear scale without curvature allows using standard spectroscopy software for further analysis. The method has been shown to work with a short focal length camera, but it is applicable to larger format, longer focal length cameras with higher spectral resolution. To the knowledge of the authors, the use of the orthographic projection for linearization of spectral dispersion has not been applied to optical spectroscopy so far. In radio interferometry however, the orthographic projection is widely used in aperture synthesis (Calabretta, 2002). The transformation equations (11) to (14) are independent of the grating, they only depend on the lens and its distortions. Therefore the distortion coefficients can be derived without grating from an astrometric analysis of an image containing a sufficiently large number of stars. This is the preferred method, as the whole field of view out to the corners contributes to the determination of the parameters. The dispersion can then be obtained from any two lines in the transformed image of a spectrum. The first results show that the method works as expected. Some improvements are still possible, such as the use of a grating with higher dispersion (first results look promising with increased resolution), or a camera with higher angular resolution and larger dynamic range. The software can also be streamlined to simplify the processing of the video files into calibrated spectra.

## Acknowledgments

This project is part of a joint effort in Meteor observation by the Swiss "Fachgruppe Meteorastronomie" (http://www.meteorastronomie.ch/), which runs a network of meteor observing stations in Switzerland and is active in promoting different aspect of meteor astronomy (path determination, fireball reporting, visual, photographic, video and radio observation). Numerous discussions within this group helped in refining this project.

## Note added in proof

A spectrum calibration method similar to (Ceplecha, 1961) can be found in (Zender et al., 2014) which includes distortion caused by an image intensifier.

*Handling Editor:* Javor Kac